\documentclass[aps,prl,twocolumn,showpacs,superscriptaddress]{revtex4-2}
\usepackage[english]{babel}
\usepackage[utf8]{inputenc}
\usepackage{amsfonts}
\usepackage{amssymb}
\usepackage[colorinlistoftodos, color=green!40, prependcaption]{todonotes}

\usepackage{amsthm}
\usepackage{mathtools}
\usepackage{physics}
\usepackage{xcolor}
\usepackage{graphicx}
\usepackage[left=23mm,right=13mm,top=35mm,columnsep=15pt]{geometry} 
\usepackage{adjustbox}
\usepackage{placeins}
\usepackage[T1]{fontenc}
\usepackage{lipsum}
\usepackage{csquotes}

\usepackage[pdftex, pdftitle={Article}, pdfauthor={Author}]{hyperref} 
\usepackage{enumitem}
\begin{document}
\title{Enhancing Quantum Entanglement Through Parametric Control of Atom-Cavity States}

\author{Arthur Vesperini}
\email[]{arthur.vesperini@unisi.it}
\affiliation{DSFTA, University of Siena, Via Roma 56, 53100 Siena, Italy}
\affiliation{INFN Sezione di Perugia, I-06123 Perugia, Italy}

\author{Roberto Franzosi}
\email[]{roberto.franzosi@unisi.it}
\affiliation{DSFTA, University of Siena, Via Roma 56, 53100 Siena, Italy}
\affiliation{QSTAR \& CNR - Istituto Nazionale di Ottica,    Largo Enrico Fermi 2, I-50125 Firenze, Italy}
\affiliation{INFN Sezione di Perugia, I-06123 Perugia, Italy}

\date{\today} 

\begin{abstract}
Dicke states form a class of entangled states that has attracted much attention for their applications in various quantum algorithms. They emerge as eigenstates of the Tavis-Cummings Hamiltonian, a simplification of the Dicke model, which describes an assembly of two-level atoms trapped in an electromagnetic cavity. In this letter, we show that in the regime where the field energy is large with respect to the atomic energy splitting, precise control of the ground state can be implemented. Specifically, pure Dicke states can be selected and produced by appropriate tuning of the parameters. This result may have important applications in quantum engineering and quantum information theory.
\end{abstract}

\keywords{Entanglement}

\maketitle

{\em Introduction.} -- 
The ability to manipulate quantum states with the aim of enhancing their entanglement degree constitutes one of the main steps to be achieved in the development of quantum technologies for the second quantum revolution. In fact, entanglement is considered an essential resource in quantum information, quantum cryptography, teleportation, and quantum computation.
Recently, quantum entanglement has also been proposed as a key resource for the realization of high-efficiency quantum batteries \cite{gyhmQuantumChargingAdvantage2022, alickiEntanglementBoostExtractable2013, binderQuantacellPowerfulCharging2015}. Indeed, long-distance entanglement allows for a charging speedup, resulting in a super-extensive power of quantum batteries. In this context, some theoretical schemes for describing quantum batteries are inspired by the Tavis-Cummings or Dicke models \cite{seahQuantumSpeedupCollisional2021,shaghaghiMicromasersQuantumBatteries2022,shaghaghiLossyMicromaserBattery2023,ferraroHighPowerCollectiveCharging2018}.
In the latter, the fast charging effect is associated with the superradiant phase transition occurring in such systems. 
Also in the context of quantum technologies and quantum computation, Dicke states constitute a valuable resource (see \cite{bartschiDeterministicPreparationDicke2019} and references therein).


In the present work, we address the study of the off-resonance Tavis-Cummings model, demonstrating that it indeed undergoes a superradiant quantum phase transition (QPT) at a finite system size, persisting as the system size approaches the thermodynamic limit. Moreover, we identify the Hamiltonian parameters that, at the quantum phase transition, drive the system's atomic ground state towards any chosen Dicke state. Among these, the most interesting one, in terms of its applications, is the maximally entangled Dicke state. This result can therefore be applied as an experimental scheme for preparing maximally entangled states.

{\em Model.} --
The Jaynes–Cummings (JC) model was initially proposed in 1963 \cite{JCpaper} to describe the interaction of a two-level atom with an electromagnetic field. Today, it is considered an intriguing model due to the various physical effects it exhibits, such as Rabi oscillations, collapses and revivals of Rabi oscillations, and the superradiant phase transition \cite{Larson_2007,PhysRevA.101.053805,PhysRevA.101.043835,PhysRevA.98.021802}.
The Tavis–Cummings (TC) model is an interesting generalization of the JC model, considering $M$ two-level atoms (qubits) interacting with a single mode of a quantized electromagnetic field \cite{PhysRev.170.379}. The TC model provides the opportunity to observe additional non-classical effects, including state squeezing and quantum state entanglement. These phenomena, in particular, are considered of primary importance, for instance, in enhancing the performance of optical atomic clocks towards the Heisenberg limit \cite{PhysRevLett.104.073602}.
The TC model Hamiltonian is given by
\begin{equation}
H = \omega_c a^\dagger a + \omega_z S_3 - \frac{\lambda}{\sqrt{M}}(a^\dagger S_- + a S_+),
\label{tcH}
\end{equation}
where $a^\dagger$ and $a$ are the creation and annihilation operators of photons in the cavity, satisfying $[a, a^\dagger] = 1$. The total spin operator components $S_j$, $j=1,2,3$, satisfy the usual commutations $[S_i, S_j] = i \epsilon_{ijk} S_k$, and $S_\pm = S_1 \pm i S_2$. Here and in the following, we consider units in which $\hbar=1$. In terms of the Pauli matrices $\sigma^\alpha_j$ ($j=1,2,3$), where $\alpha=0,\ldots,M-1$ runs over the index of the distinguishable atoms, we have $S_i = \sum^{M-1}_{\alpha=0} \sigma^\alpha_i/2$.
The model is characterized by three tuning parameters: the photon frequency, denoted as $\omega_c$, the atomic energy splitting, denoted as $\omega_z$, and the photon–atom coupling strength, denoted as $\lambda$. The TC model possesses desirable properties as it is fully solvable \cite{PhysRev.170.379}.
In fact, the calculation of the eigenstates can be simplified by expressing the Hamiltonian of the full system as a sum of two commuting terms, namely $H = H^I + H^{II}$ and, where
$H^{I}=\omega_c (a^\dagger a+ S_3)$, $H^{II}= \omega_c (\eta-1) S_3 - \omega_c g\sqrt{\eta / M} (a^\dagger 
S_- + a S_+)$, where $\eta = \omega_z /\omega_c$ and $g=\lambda/\sqrt{\omega_c \omega_z} $. 

For the sake of simplicity, from now on we will set $\omega_c=1$ or, in other words, express energies in units of $\omega_c$. The model can thus be studied by solely considering the two parameters $\eta$ and $g$, rewriting the Hamiltonian as
\begin{equation}
H = a^\dagger a + \eta S_3 - \frac{g\sqrt{\eta}}{\sqrt{M}}(a^\dagger S_- + a S_+) \, .
\label{tcH_new}
\end{equation}

{\em Quantum phase transition.} --
The TC model undergoes a quantum phase transition, the nature of which can be understood by studying the system's ground state as a function of the Hamiltonian parameters.
While the TC model has an infinite-dimensional Hilbert space, since $[H^I, H^{II}]=0$, the energy spectrum can be organized into multiplets of simultaneous eigenstates for $H$, $H^I$, and $H^{II}$ \cite{PhysRev.170.379}. Each multiplet is an eigenspace of $H^I$ that is identified by an index $k\in \mathbb{N}$. 
In particular, the eigenvalues of $H^I$ and their corresponding multiplicities are $E^I_k = k-M/2$ and $d^I_k = \min(k+1,M+1)$ for $k\in \mathbb{N}$, respectively. 
The energy spectrum of the full Hamiltonian is therefore derived by diagonalizing $H^{II}$ in each eigenspace ${\cal H}^I_k$ (of dimension $d^I_k$), for $k \in \mathbb{N}$. We denote by $|n, M_3\rangle$ the tensor product of an $n$-photons Fock state and a normalized eigenstate of $S_3$ with eigenvalue $M_3$, for $-M/2 \leq M_3\leq M/2$. The $k$-th eigenspace ${\cal H}^I_k$ is spanned by the vector basis $|k,-M/2\rangle, \ldots, |0,k-M/2 \rangle$, for $k\leq M$, and $|k,-M/2\rangle, \ldots, |k-M,M/2 \rangle$, for $k>M$. Note
that in each eigenspace ${\cal H}^I_k$ we have $k=n_\varphi + M_3 + M/2$, where $n_\varphi \geq 0$ is the number of photons in the cavity. The general form of the eigenstates of $H$ in ${\cal H}^I_k$ is
\begin{equation}
|\Psi\rangle = \!\!\!\!\sum_{n=0}^{\min\{k,M\}}\!\!\!\!a_n|k-n, n - M/2\rangle,\; \textrm{with}\; k\in\mathbb{N} \, .
\label{generic_eistate}
\end{equation}
Let us denote by $E_k(\eta, g)$ the lowest eigenvalue of $H$ in each eigenspace ${\cal H}^I_k$, depending on the Hamiltonian parameters. The corresponding ground state for $H$ is
\begin{equation}
|GS (\eta,g) \rangle  =  |E_{k^*}(\eta,g) \rangle  
\end{equation}
where $k^*\in\mathbb{N}$ is such that
\begin{equation}
E_{k^*} (\eta,g)= \min_k 
\{ E_k (\eta,g) \}\, ,
\end{equation}
thus $k^* = k^* (\eta,g)$ is named excitation number of ground state of the full Hamiltonian.

For $g<1$, $|GS (\eta,g)\rangle=|0,-M/2\rangle$ and $k^*=0$.
However, at $g=g_1:=1$, the energy level $E_0(\eta,g)$ crosses the level $E_1(\eta,g)$. Corresponding to this level crossing $k^*$ jumps from $0$ to $1$. This first level crossing is followed by further crossings between the minimum energy levels of successive multiplets as the magnitude of $g$ increases. 
If, by increasing the magnitude of $g$, $E_k(\eta,g)$, the minimum energy level of $H$, crosses the minimum energy level of $H$ $E_{k+1}(\eta,g)$ at $g=g_{k+1}(\eta)$, then at this value of $g$, the value of $k^*$ changes from $k$ to $k+1$.
Since $n_\varphi = k - M_3  -M/2\geq k - M$, the greater $g$ is, the greater $k^*$ becomes, and consequently, the greater $n_\varphi$ becomes.
Remarkably, in the limit of strong spin energy separation, i.e., as $\eta$ approaches infinity, all the crossing-level points converge at the quantum phase transition point $g=1$. We expect that, in the opposite limit, as $\eta$ approaches zero, the differences between successive critical values $g_k$ increase. We will show below that it is the case, provided that $g$ remains small enough.

{\em Maximally entangled Dicke states.} -- 
The eigenstates $|S=M/2,M_3=n-M/2\rangle=|D^M_n\rangle$ of $S_3$ are in fact Dicke states, namely degenerate states of $n$ excited qubits. They are defined as
\begin{equation}
|D^{M}_n\rangle =\binom{M}{n}^{-1/2}
\sum_j P_j\{|e\rangle^{\otimes n} \otimes |g\rangle^{\otimes (M-n)}\} \, ,
\label{dickes}
\end{equation}
where we denote with $\sum_j P_j$ the sum over all the possible permutations and, for $\mu=0,\ldots,M-1$, $|e\rangle^\mu$ and $|g\rangle^\mu$ are the eigenstates of $\sigma^\mu_3$ with eigenvalues $+1$ and $-1$, respectively. 

Yet, the Dicke states are known to possess a number of interesting properties. In particular, some of them are known to exhibit \textit{genuine multipartite entanglement}. It is thus of great interest to devise ways to generate such target states in a controlled fashion. 

Hereafter, we use the following definition of pure state entanglement per qubit, derived by the Entanglement Distance (ED) \cite{PhysRevA.101.042129},
\begin{equation}
E\big(| \psi\rangle\big):=1 - \frac{1}{M}\sum^{M-1}_{\mu=0}\sum^3_{j=1} \big|\langle \psi|\sigma_j^\mu| \psi\rangle\big|^2,
\end{equation}
where the first sum runs over all the qubits.

For even $M$, the state $|D^{M}_{M/2}\rangle$ maximizes the ED $E$, since it is $E\big(|D^{M}_{M/2}\rangle\big)=1$. 
On the other hand, for odd $M$, the most entangled Dicke states are $|D^{M}_{M/2\pm1}\rangle$, with $E\big(|D^{M}_{M/2\pm1}\rangle\big)=1-4/M^2$.

{\em Parametric state control.} -- 
The form of Hamiltonian \eqref{tcH_new} suggests that $\eta$ is a key parameter governing the probability amplitudes $a_n$. In fact, the smaller $\eta$ is, the greater the number of excited atoms, since a smaller $\eta$ corresponds to a smaller energy cost of the atomic excitation compared to that of photons.
Guided by this observation, we expect that straightforward control of the ground state can be implemented through fine-tuning the parameters $(g,\eta)$,  given the dependence $k^*=k^*(g,\eta)$ and $a_n=a_n(g,\eta)$. 

In particular, the atomic part of the ground state writes
\begin{equation}\label{rhos}
\begin{split}
\rho_s(g,\eta)&=\Tr_\varphi\big[|\Psi\rangle\langle\Psi|\big]\\
&= \sum_{n=0}^{\min\{k,M\}}w_n\left(g,\eta\right)|D^{M}_n\rangle\langle D^{M}_n|,
\end{split}
\end{equation}
where $\Tr_\varphi\left[\cdot\right]$ denotes the partial trace over the photonic part of the system, and $w_n=a_n^2$. Thus, controlling the weights $w_n(g,\eta)$, one may be able to engineer mixtures of Dicke states. 

Obviously, for all practical purposes, the most interesting targets are pure Dicke states and, even more, the maximally entangled ones.

{\em Perturbative development.} -- 
Restricting ourselves to an eigenspace ${\cal H}^I_k$, we can rewrite the Hamiltonian as $H_k = H^I_k + H^{II}_k$, where $H^I_k= E^I_k\mathbb{I}$, and
\begin{equation}\label{perturb_H}
H^{II}_k = H_k^{(0)} + \sqrt{\eta}H_k^{(1)} + \eta H_k^{(2)}\, ,
\end{equation}
with $H_k^{(0)}=-S_3$, $H_k^{(1)}=-g/\sqrt{M} (a^\dagger S_- + a S_+)$ and $H_k^{(2)}=S_3$. We set aside the constant term $H^I_k$ for now.\\
Considering $\sqrt{\eta}\ll 1$, we can apply perturbation theory to $H^{II}_k$. Namely, we assume that its ground state and ground state energy can be perturbatively expanded in the parameter $\sqrt{\eta}$, that is $\ket{ E_k}=\sum_j \eta^{j/2}| E_k^{(j)}\rangle$ and $E_k^{II} =\sum_j \eta^{j/2}E_k^{(j)}$, where $j=0,1,2,\cdots$

\begin{figure}[htb!]
    \centering
    \includegraphics[width=0.99\columnwidth]{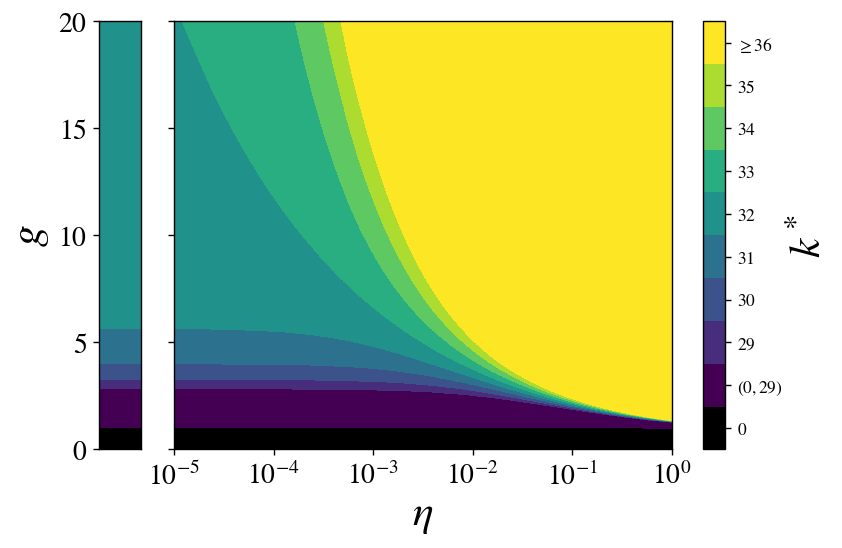}
    \includegraphics[width=0.99\columnwidth]{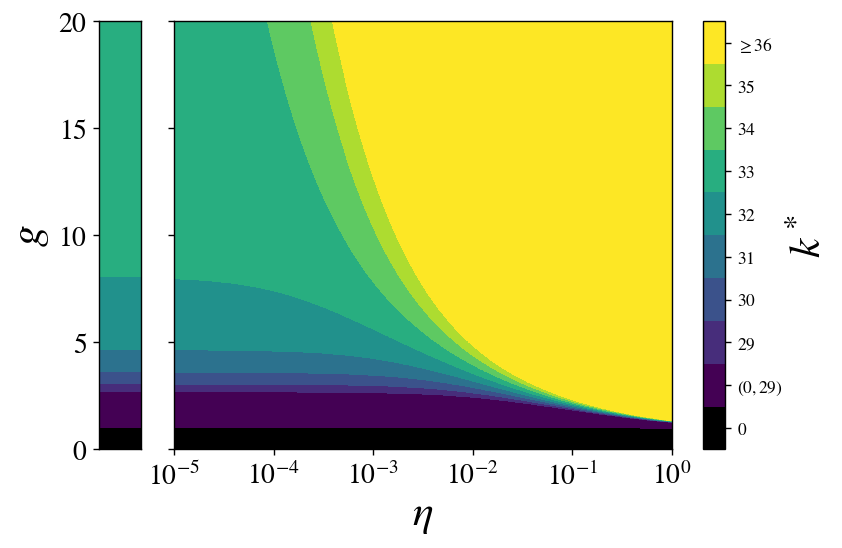}
    \caption{The figures show the value of $k^*$ as a function of $\eta$ and $g$, for $M=64$ (upper figure) and for $M=65$ (lower figure). We emphasize that, for $\eta\ll 1$ (left-hand side of the figures), $k^*$ reaches visible plateaus while, for higher values of $\eta$ (right-hand side of the figures), it increases very rapidly, accounting for superradiance. For the sake of clarity, we chose to limit the accuracy of the $ k^*$ level, only representing a few different values of interest, from $k^*=N/2-3$ to $k^*=N/2+3$. The left coloured band shows the expected results in the asymptotic regime $\eta\to 0$, retrieved in our perturbative analysis.}
    \label{fig:k_vs_eta-g}
\end{figure}

From the eigenvalue equation $H^{II}_k \ket{ E_k} = E_k\ket{ E_k}$, we straightforwardly obtain, after some computations, solutions up to second order in $\sqrt{\eta}$. 
At zeroth order, we have $ E_k^{(0)} = -(k - M/2)$ and $ |E_k^{(0)}\rangle=|0, k-M/2\rangle$ for $k\leq M$ ($ |E_k^{(0)}\rangle=|k-M, M/2\rangle$ for $k> M$).
At first order, we have $E_k^{(1)} =0 $ and $| E_k^{(1)}\rangle = g\sqrt{k\left(1 - \frac{k-1}{M}\right)} | 1, k-1 -M/2 \rangle$ for $k\leq M$ ($| E_k^{(1)}\rangle = g | k-M + 1, M/2 -1 \rangle$ for $k> M$).
Finally, the second order energy correction reads $E_k^{(2)} = - g^2k\left(1-\frac{k-1}{M}\right) + (k- M/2)$ for $k\leq M$ ($E_k^{(2)} = - g^2 + M/2$ for $k> M$).
It results the total ground state energy $E_k = E_k^I + E_k^{II}$ in the $k$-th eigenspace
\begin{equation}
\begin{split}
    E_k &= \eta\left( k - \dfrac{M}{2} - g^2k\left(1-\frac{k-1}{M}\right)\right) + o(\eta^{3/2}) \, , \ k\leq M \\
    E_k &= k - M + \eta\left(\dfrac{M}{2} - g^2 \right) + o(\eta^{3/2})\, , \ k> M \, .
\end{split}
\end{equation}
Solving $E_{k+1}-E_k=0$, we retrieve the position of the level crossings.
In the case $k> M$ there are no solutions, in the case $k \leq M$ we get
\begin{equation}
g_k\approx\sqrt{\frac{M}{M - 2k + 2}}\, ,
\end{equation}
and, therefore, for the total excitation number of the ground state of the full Hamiltonian as a function of $g$ we have
\begin{equation}
k^*\approx \left\lceil\frac{M}{2}\left( 1 -  1/g^2\right)\right\rceil\, .
\end{equation}
\begin{figure}[htb!]
    \centering
    \includegraphics[width=0.99\columnwidth]{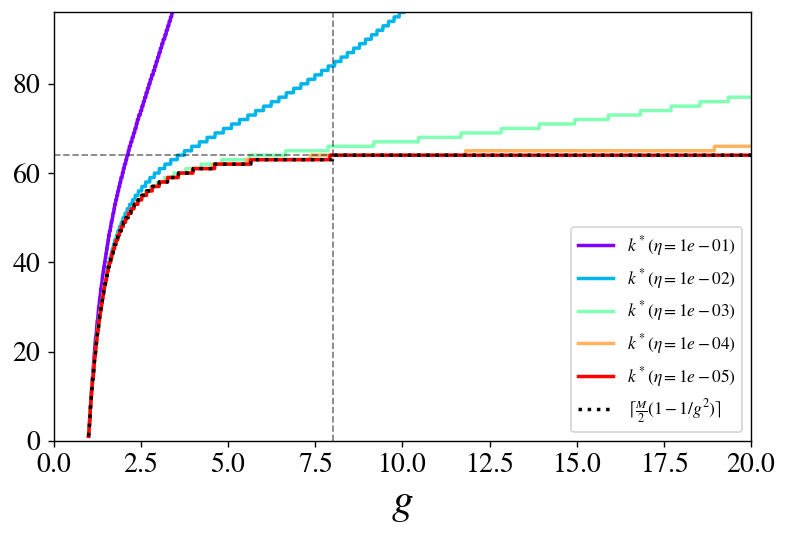}
    \caption{Total number of excitations in the ground state of a system with $N=64$ atoms, as computed by our perturbative development (black dotted line), and retrieved numerically for different values of the parameter $\eta$ (colored plain lines). The vertical dashed line marks the position of $g=\sqrt{M/2}$, expected onset of the $k^*=M/2$ plateau, while the horizontal dashed line spots the value $M/2=32$.}
    \label{fig:perturb_fit}
\end{figure}
Clearly, in the perturbative regime, the targeted value $k^*=M/2$ ($k^*=(M+1)/2$) is reached for high enough values of $g$, namely when $g>\sqrt{M/2}$ ($g>\sqrt{M}$) for $M$ even (odd). No further level crossing is predicted by our perturbative analysis; however, for any finite $\eta$, this analysis ceases to be valid for large $g$. This is because the term $\sqrt{\eta}H_k^{(1)}$ scales as $g\sqrt{\eta}$. An appropriate trade-off should thus be found, with high enough $g$ and low enough $g\sqrt{\eta}$.
In Figures \ref{fig:perturb_fit} and \ref{fig:k_vs_eta-g}, we see that the agreement with numerical results is quite good in the regime where $\eta \ll 1$ and $g\sqrt{\eta}\ll 1$. 
Furthermore, the perturbative expansion predicts that the atomic part of the ground state is, up to a correction of order $o(\eta g^2 k)$, in the pure state $|D^M_k\rangle\langle D^M_k|$.

\begin{figure}[htb!]
    \centering
    \includegraphics[width=0.99\columnwidth]{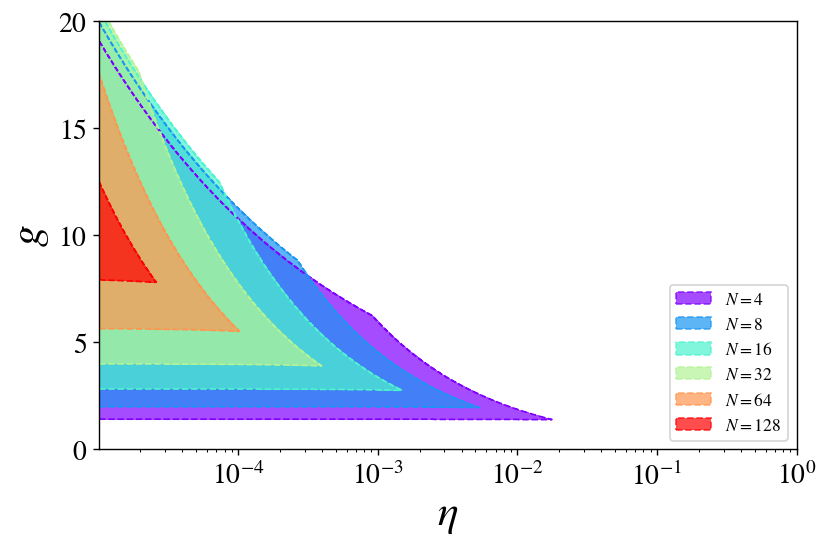}
    \includegraphics[width=0.99\columnwidth]{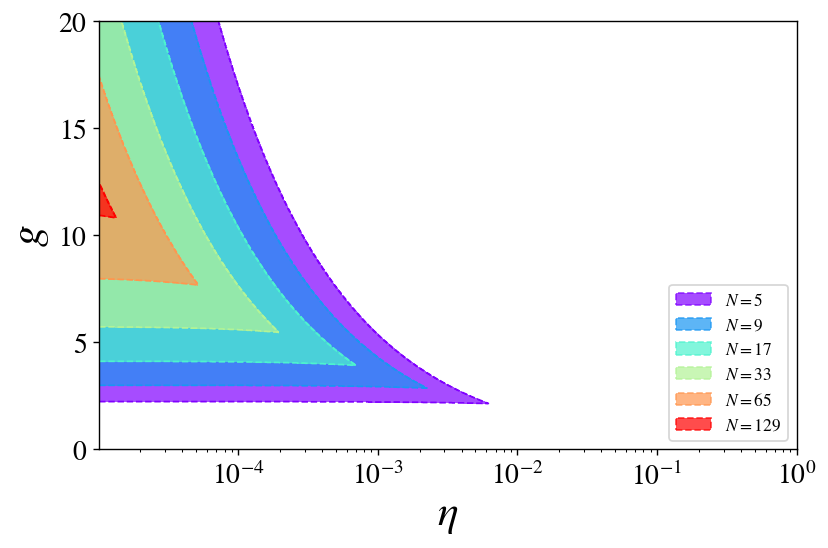}
    \caption{The figure shows, for different system size $M$, the parametric region such that $w_{_{M/2}}\geq 0.95$, for even $M$ (upper figure) and $w_{_{M/2+1}}\geq 0.95$ for odd $M$ (lower figure).}
    \label{fig:all-N-bool}
\end{figure}

We numerically computed $w_{M/2}\left(g,\eta\right)$ ($w_{M/2+1}\left(g,\eta\right)$) for a wide range of values of these parameters, and different even (odd) $M$, as shown in Fig. \ref{fig:all-N-bool}. 
As expected, for any \textit{finite} $M$, there exists a parametric region for which $w_{M/2}\approx1$, and $$\forall M<\infty,\;\exists\;\Tilde{g} \text{ such that } w_{M/2}\left(g=\Tilde{g},\eta\right)\underset{\eta\to 0}{\longrightarrow} 1.$$ 

It results that the fine-tuning of the parameters of the TC model allows for the generation of an almost pure highly entangled state of $M$ qubits. 

Evidently, as $M$ increases, the parametric region of interest shrinks (as can be seen in Fig. \ref{fig:all-N-bool}); this implies that such fine-tuning cannot be achieved for arbitrary system sizes.

It is worth noting that a mere projective measurement on the photonic degrees of freedom of the ground state suffices to obtain a pure Dicke state.
To generate the state $|D^M_k\rangle$, we simply place ourselves in the bulk of the parametric region where $k^*=k$, cool the system down to its ground state, and perform a measurement of the photon number. The parameters can be fine-tuned so that the latter has a high probability to be null, namely $w_k\approx 1$, as can be seen in Figure \ref{fig:all-N-bool} in the specific case $k=M/2$; in this case, the atomic part of the system is in the desired pure state. It has on the other hand a probability of order $o(\eta g^2 k)$ to be non-zero; in this case, we reinitialize the process. 

We showed in this letter that a chosen pure Dicke state may be obtained as the atomic part of the ground state of the Tavis-Cummings model, upon appropriate tuning of the parameters. Furthermore, the maximally entangled state $|D_{M/2}^M\rangle$, most useful in algorithmic applications, is shown to be particularly easy to target, as it corresponds to a wide parametric region. 
Follow-ups of this work include the search of possible physical implementations, and to take into account the noise occurring at finite temperature and in opened systems. One could further study to which extent these results are applicable to other models, such as the Dicke model.

This framework could be of great use in quantum technologies.

{\em Acknowledgments.} --
We acknowledge support from the RESEARCH SUPPORT PLAN 2022 - Call for applications for funding allocation to research projects curiosity-driven (F CUR) - Project ”Entanglement Protection of Qubits’ Dynamics in a Cavity”– EPQDC and from INFN-Pisa.
Furthermore, 
we acknowledge support from the Italian National Group of Mathematical Physics (GNFM-INdAM).

\bibliography{references}

\appendix*

\end{document}